\documentclass[a4paper,fleqn,usenatbib,useAMS]{mnras}

\usepackage{graphicx}	
\usepackage{amsmath}	
\usepackage{amssymb}	
\usepackage{multicol}        
\usepackage{bm}		
\usepackage{pdflscape}	
\usepackage[encapsulated]{CJK}
\usepackage{ucs}
\usepackage[utf8x]{inputenc}
\usepackage{multirow}
\usepackage{booktabs,caption}
\usepackage[flushleft]{threeparttable}

\usepackage[T1]{fontenc}
\usepackage{ae,aecompl}

\usepackage{newtxtext,newtxmath}

\title[IPHASX J191104.8+060845]{Detailed studies of IPHAS sources - III. The highly extinguished bipolar planetary nebula IPHASX J191104.8+060845}
\author[J.B.\,Rodr\'{i}guez-Gonz\'{a}lez et al.]{J.B.\,Rodr\'{i}guez-Gonz\'{a}lez$^{1}$\thanks{E-mail:j.rodriguez@irya.unam.mx}, L.\,Sabin$^{2}$, J.A.\,Toal\'{a}$^{1}$, S. Zavala$^{3}$, G.\,Ramos-Larios$^{4}$,\newauthor M.A.\,Guerrero$^{5}$, Q.A. Parker$^{6,7}$, P.F.\,Guill\'{e}n$^{8}$ and A. Ritter$^{6}$ \\
  $^{1}$Instituto de Radioastronom\'{i}a y Astrof\'{i}sica (IRyA), UNAM Campus Morelia, Apartado postal 3-72, 58090 Morelia, Michoacan, Mexico\\
  $^{2}$Instituto de Astronom\'{i}a,  Universidad Nacional Aut\'{o}noma de M\'{e}xico, Apdo. Postal 877, 22800 Ensenada, B.C., Mexico\\
  $^{3}$Tecnol\'ogico Nacional de M\'exico / I. T. Ensenada (TecNM/ITE), Blvd. Tecnol\'ogico No. 150, C. P. 22780, Ensenada, B. C., Mexico\\
  $^{4}$Instituto de Astronom\'{i}a y Meteorolog\'{i}a, Universidad de Guadalajara, Av. Vallarta 2602, Arcos Vallarta, 44130 Guadalajara, Mexico\\
  $^{5}$Instituto de Astrof\'{i}sica de Andaluc\'{i}a (IAA-CSIC), Glorieta de la Astronom\'{i}a S/N, E-18008 Granada, Spain\\
  $^{6}$Physics Department, CYM Building, The University of Hong Kong, Pokfulam, Hong Kong SAR, China.\\
 $^{7}$Laboratory for Space Research, Hong Kong University, Hong Kong, China\\
  $^{8}$Instituto de Astronom\'{i}a,  Universidad Nacional Aut\'{o}noma de México, Observatorio Astron\'{o}mico Nacional, Ensenada, Baja California, Mexico
}

\pubyear{2020}

\begin{document}
\label{firstpage}
\pagerange{\pageref{firstpage}--\pageref{lastpage}}
\maketitle

\begin{abstract}
  
We present the first detailed study of the bipolar planetary nebula (PN)
IPHASX J191104.8+060845 (PN G\,040.6$-$01.5) discovered as part of the Isaac
Newton Telescope Photometric H$\alpha$ Survey of the Northern Galactic
plane (IPHAS). We present Nordic Optical Telescope (NOT) narrow-band images to
unveil its true morphology. This PN consists of a main cavity
with two newly uncovered extended low-surface brightness lobes located towards the NW and SE
directions. Using near-IR {\it WISE} images we unveiled the presence
of a barrel like structure, which surrounds the main cavity, which would 
explain the dark lane towards the equatorial regions. We also use
Gran Telescopio de Canarias (GTC) spectra to study the physical
properties of this PN. We emphasise the potential of old PNe detected
in IPHAS to study the final stages of the evolution of the circumstellar medium around solar-like stars.
 
\end{abstract}

\begin{keywords}
stars: evolution --- stars: winds, outflows --- planetary nebulae:
general --- planetary nebulae: individual: IPHASX\,J191104.8+060845
\end{keywords}




\section{INTRODUCTION}
\label{sec:intro}

Planetary nebulae (PNe) are the ejected remnants of the last
evolutionary stages of low- and intermediate-mass stars
($M_\mathrm{ZAMS}\lesssim$~1--8~M$_{\odot}$). During the asymptotic
giant branch (AGB) phase, these stars lose mass through a slow
($v_\mathrm{AGB}\approx$10--30~km~s$^{-1}$) and dense
($\dot{M}\lesssim10^{-5}$~M$_{\odot}$~yr$^{-1}$) wind that produces
shells that interact directly with the circumstellar medium
\cite[see][]{Tweedy1994}. Stars in the AGB phase exhibit a relatively
low effective temperature of $T_\mathrm{eff}\approx3,000-6,000$~K
which allows the formation of dust shells \citep[][]{Cox2012}. When
the star finally expels its outer layers, a hot stellar core is
exposed. These post-AGB stars can produce fast winds
\citep[$v_{\infty}$= 500 -- 4,000~km~s$^{-1}$;][]{Guerrero2013} that
will shock and compress the AGB material whilst it is being
photoionised by the stellar raising UV flux. The combination of such
effects produces a PN now emanating from the heating stellar core.

PNe are the outcome of the most numerous population of
stars in the Universe (e.g. solar-like stars). In our Galaxy, the number of PNe has been
estimated to range between 6,600 -- 46,000 depending on the adopted
scenarios and nebular sizes
\citep[e.g.,][]{DeMarco2005,Frew2006,Moe2006,Zijlstra1991}. Around 3,500 PNe have been 
confirmed in our Galaxy thanks principally to the advent of wide field narrow band 
H$\alpha$ surveys of the Galactic Plane \citep[e.g. ][]{Parker2005, Drew2005} that 
permitted discovery of $\sim$2,000 additional PNe 
\citep[e.g. ][]{Parker2006,Miszalski2008,Sabin2014} and now increasingly thanks to the 
tireless work of the amateur community \citep[e.g.,][]{Kronberger2014}. A full PNe 
registry can be found in the Hong Kong/AAO/Strasbourg H$\alpha$ (HASH) PN 
database \citep{Parker2016}. Increasing the number of known PNe is of 
importance to improve our understanding of the luminosity function of PNe, 
galactic chemical enrichment, stellar population synthesis, and galactic 
abundance gradients.

\begin{figure*}
\begin{center}
  \includegraphics[angle=0,width=0.95\linewidth]{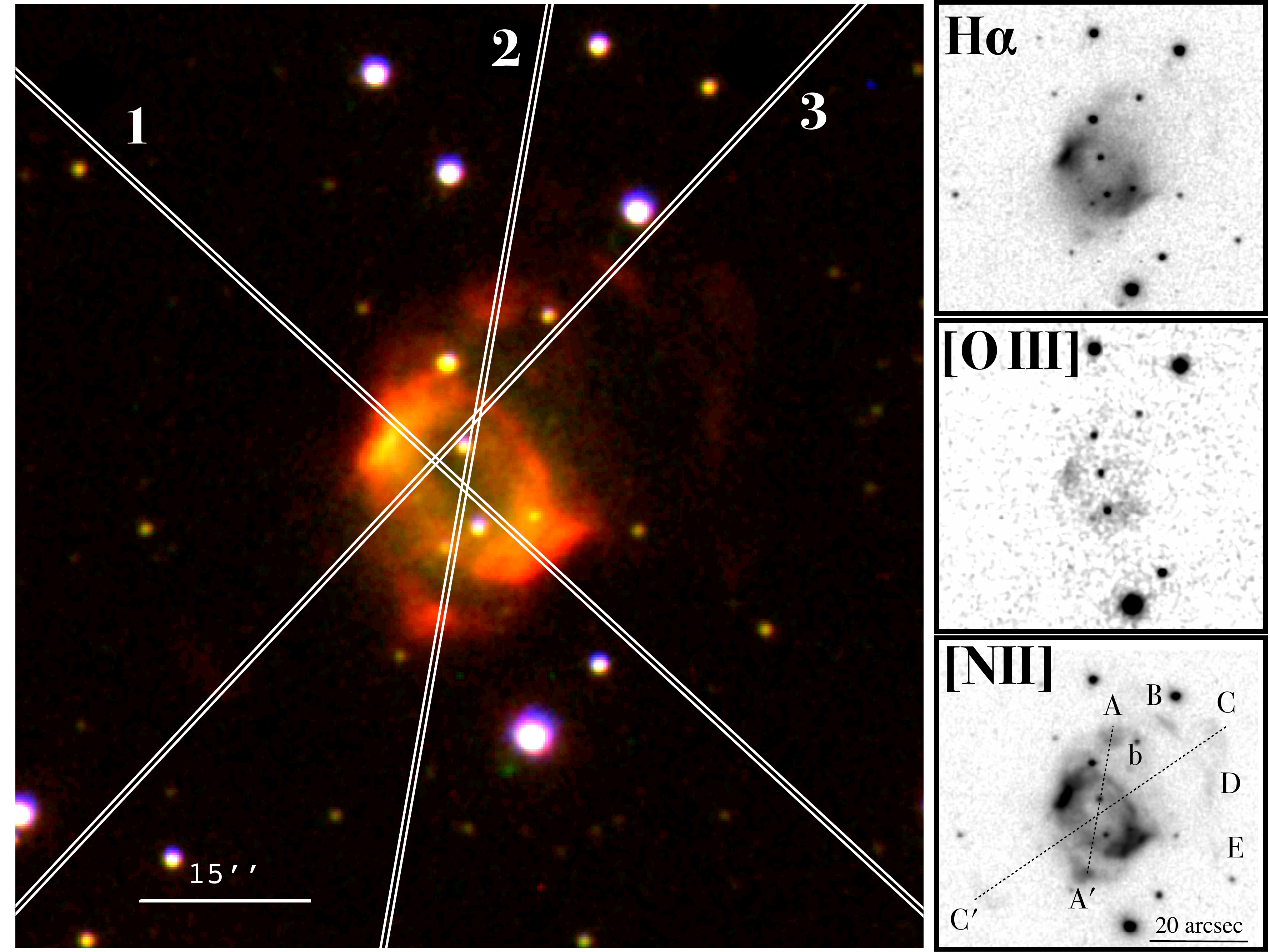}
\caption{Colour-composite picture (left) and narrow band images
  (right) of IPHASX\, J191104.8+060845 obtained with the Nordic
  Optical Telescope. Red, green and blue correspond to [N\,{\sc ii}],
  H$\alpha$ and [O\,{\sc iii}], respectively. The position of the MES
  Echelle slits are labeled as 1, 2 and 3 and correspond to PA of
  $+47^\circ$, $-10^{\circ}$ and $-43^\circ$, respectively. The clumps
  around the main nebular shell are labeled in the [N\,{\sc ii}] image
  (lower right panel) referred in Section~3.2. The dotted lines
  connect the lobes of what we believe to be pairs of bipolar
  ejections. In all panels north is up, east to the left.}
\label{NOT}
\end{center}
\end{figure*}

The Isaac Newton Telescope Photometric H$\alpha$ Survey of the
Northern Galactic plane
\citep[IPHAS;][]{Drew2005,Gonzalez2008,Barentsen2014} has been
designed to unveil the presence of mostly faint emission-line objects
located in the Northern Galactic plane and hence characterised by
higher extinction and by their advanced evolutionary phase or large
distances \citep{Sabin2014}. IPHAS is a photometric CCD survey which
covered the inner regions of the northern plane ($-5^\circ < b <
5^\circ$ and $29^\circ < l < 215^\circ$) with a total area coverage of
1800~deg$^{2}$. The survey made used of the Wide Field Camera (WFC) on
the Isaac Newton Telescope (INT) 2.5~m at La Palma in the Canary
Islands (Spain) which has a field of view of $34\times34$~arcmin$^{2}$
and a pixel scale of 0.33~arcsec~pix$^{-1}$. The median seeing value
was 1.1~arcsec. The good spatial resolution of the IPHAS data, as well
as the use of binned mosaics \citep{Sabin2008}, made this survey ideal
for the detection of extended evolved PNe at the faint end of the PNe
luminosity function. As a result, a total of 159 likely and possible
PNe were confirmed with all types of morphologies
\citep[see][]{Sabin2014}. An additional sample of PNe will be
presented soon (Ritter et al. in prep.).  In order to confirm the PN
nature of objects detected in IPHAS, a combination of different
observational tools such as deep images and spectra is required
\citep[e.g.,][]{Frew2010}. This is the third paper of a series in
which we characterise in detail interesting PNe from IPHAS (Sabin et
al. 2020 and Guerrero et al 2020, both submitted to MNRAS).  Here we
present the analysis of the PN IPHASXJ\,191104.8+060845 (a.k.a. PN
G\,040.6$-$01.5; hereafter J191104), confirmed spectroscopically on
the GTC/OSIRIS (2018 May 17). We characterise its nature, morphology
and physical properties.  The paper is organized as follows. In
Section~2 we present our observations, in Section~3 we present the
analysis of our images and spectra and in Section~4 we discuss our
results. Finally, the summary is presented in Section~5.

\begin{figure*}
\begin{center}
  \includegraphics[angle=0,width=1\linewidth]{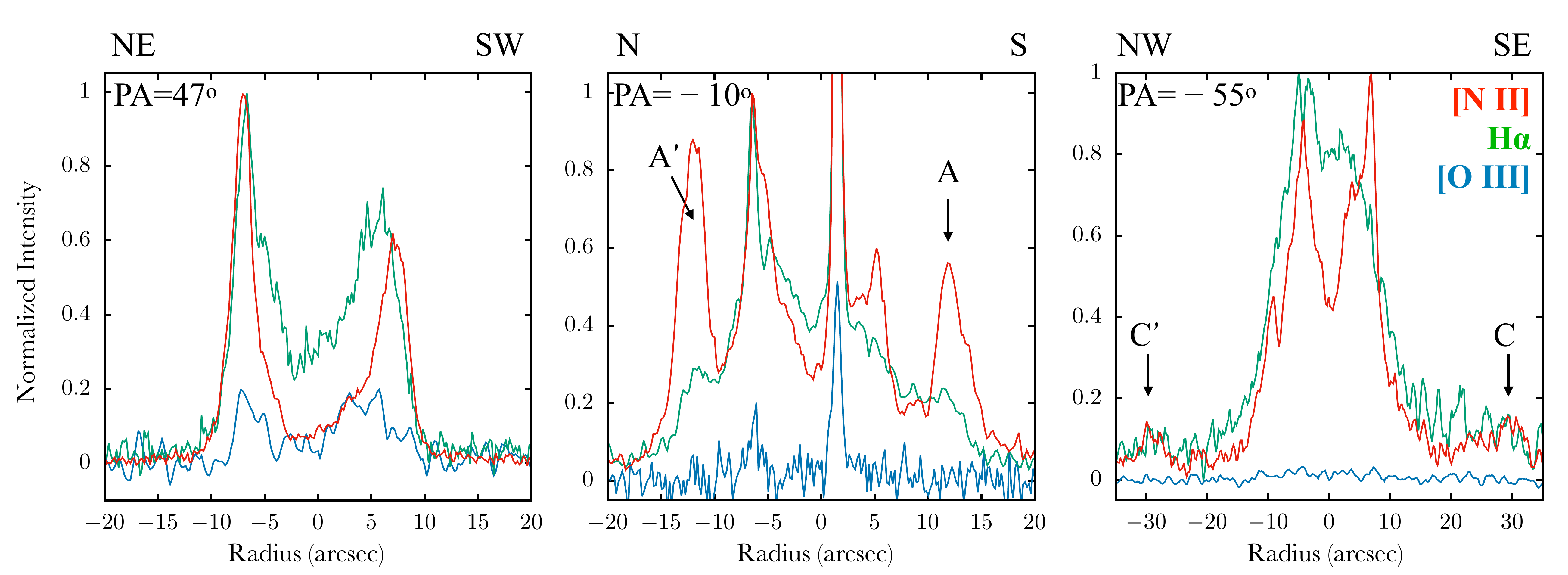}
\caption{Normalised surface brightness profiles extracted from the NOT 
images. The left, center and right panels 
show the corresponding profiles extracted from PA=$+47^{\circ}$, 
$-10^{\circ}$ and $-55^{\circ}$, respectively. The contribution 
from a star in the field of view of the three emission lines 
can be seen at $\sim2''$ in the middle panel.}
\label{fig:Sab2_profiles}
\end{center}
\end{figure*}

\section{Observations}

\subsection{ALFOSC Optical Imaging}

We observed J191104 with the Nordic Optical Telescope (NOT) at the
Observatorio de El Roque de los Muchachos (ORM) in La Palma (Spain)
with the Alhambra Faint Object Spectrograph and Camera
(ALFOSC)\footnote{\url{http://www.not.iac.es/instruments/alfosc/}} on
2018 June 14 using H$\alpha$, [O\,{\sc iii}] and [N\,{\sc ii}]
narrow-band filters.  The central wavelength and FWHM are 6563~\AA\,
and 10~\AA\, for H$\alpha$, 5007~\AA\, and 30~\AA\, for [O\,{\sc iii}]
and 6584~\AA\, and 10~\AA\, for [N\,{\sc ii}].  Bias and flat-field
images were obtained to produce the final narrow-band filter images
following standard procedures in {\sc iraf} \citep{Tody1993}.
Individual H$\alpha$, [N\,{\sc ii}] and [O\,{\sc iii}] narrow-band
filter images as well as a colour-composite image of J191104 are
presented in Figure~\ref{NOT}.

\subsection{OSIRIS Long slit spectroscopy}

Long slit intermediate resolution spectra were acquired with the
Optical System for Imaging and low-Intermediate Resolution Integrated
Spectroscopy instrument \citep[OSIRIS;][]{Cepa2003} mounted on the
10.4~m GTC telescope at the ORM. The observations were performed on
2018 May 17 using the two Marconi CCD42-82 (2048$\times$4096)
detectors with a 2$\times$2 binning leading to a spatial scale of
0$\farcs$254\,pix$^{-1}$. We used the R1000B grism covering a spectral
range from 3630\AA\, to 7500\AA\, with a spectral dispersion of
2.12\,\AA\,pix$^{-1}$. The slit width was set to 0.8\arcsec and the
7.4\arcmin\ length slit was placed at a position angle (PA)
+47$^\circ$ through the main elongated bright structure. We obtained
two spectra of 1000s exposure each. The HgAr and Ne lamps were used
for wavelength calibration and the spectrophotometric standard star GD
140 for flux calibration. The observations were performed during dark
time and clear sky but the seeing conditio of 1.7\arcsec were not
optimal. Data reduction was performed using standard {\sc iraf}
routines.

\subsection{MES long slit high resolution spectroscopy }

We obtained long-slit high dispersion optical spectra of J191104 with
the Manchester Echelle Spectrograph (MES; \citealt{Meaburn2003})
mounted on the 2.12~m telescope at the Observatorio Astron\'{o}mico
Nacional, San Pedro M\'{a}rtir (OAN-SPM, Mexico). The data were
obtained on 2018 July 22 and 2019 June 27 with a 2048$\times$2048
pixels E2V CCD detector with pixel size of 13.5
$\mu$m\,pixel$^{-1}$. MES has a fixed slit length of 6$\farcm$5 and we
set the slit width to 150 $\mu$m corresponding to
$\sim$1$\farcs$9. Two different pixel binning were used, 2$\times$2
and 4$\times$4, leading to spatial scales of
0$\farcs$351\,pixel$^{-1}$ and 0$\farcs$702 \,pixel$^{-1}$,
respectively. The exposures times were 1200~s and 1800~s and three
spectra were taken with the H$\alpha$ filter with $\Delta\lambda$ = 90
\AA\ to isolate the 87th order (0.05 and 0.1 \AA\,pixel$^{-1}$
spectral scale for the 2$\times$2 binning and 4$\times$4 binning,
respectively). The slits were arranged at PAs $+47^\circ$,
$-10^\circ$, $-43^\circ$ in order to cover the main morphological
structures of the target. The slit positions are shown in
Figure~\ref{NOT} and labeled as Slit\,1, 2 and 3, respectively. The
spectral range includes the H$\alpha$ and [N\,{\sc ii}]
$\lambda\lambda$ 6548,6584\AA\, emission lines.

\section{Results}

\subsection{Morphology}

The NOT images presented in Figure~\ref{NOT} show that J191104 is a
nebula with a clear bipolar morphology. We thus re-classify the
morphology as Bspam following the adopted HASH scheme
\citep[see][]{Parker2006}. Its main nebular shell exhibits a bright
ellipse detected in [N\,{\sc ii}] with lesser contribution from
H$\alpha$ and marginally detected in [O\,{\sc iii}]. This structure
appears brighter at the south-west and north-east and is
18\arcsec$\times$12\arcsec in size. Assuming that it corresponds to a
projected circular ring, we estimate that it is tilted 42$^{\circ}$
with respect to the plane of the sky.

A collection of knots detected in the [N\,{\sc ii}] image can be seen
towards the north-west from the central ring. These have been labeled
from A to E (see lower-right panel of Fig.~\ref{NOT}). Only two of
these seem to have counterparts towards the south-east from the
central ring. We have labeled these as A$'$ and C$'$. The pairs
A--A$'$ and C--C$'$, connected with dashed-lines in the bottom-right
panel of Figure ~\ref{NOT}, are equidistant from the center of the
ring at $\sim$14$''$ and $\sim$30$''$. The pair A--A$'$ is also
detected in the H$\alpha$ narrow-band image but the position of the
other knots is only suggested in this image. None of these structures
are detected in the [O\,{\sc iii}] image.
 
 \begin{figure*}
\begin{center}
  \includegraphics[angle=0,width=0.65\linewidth]{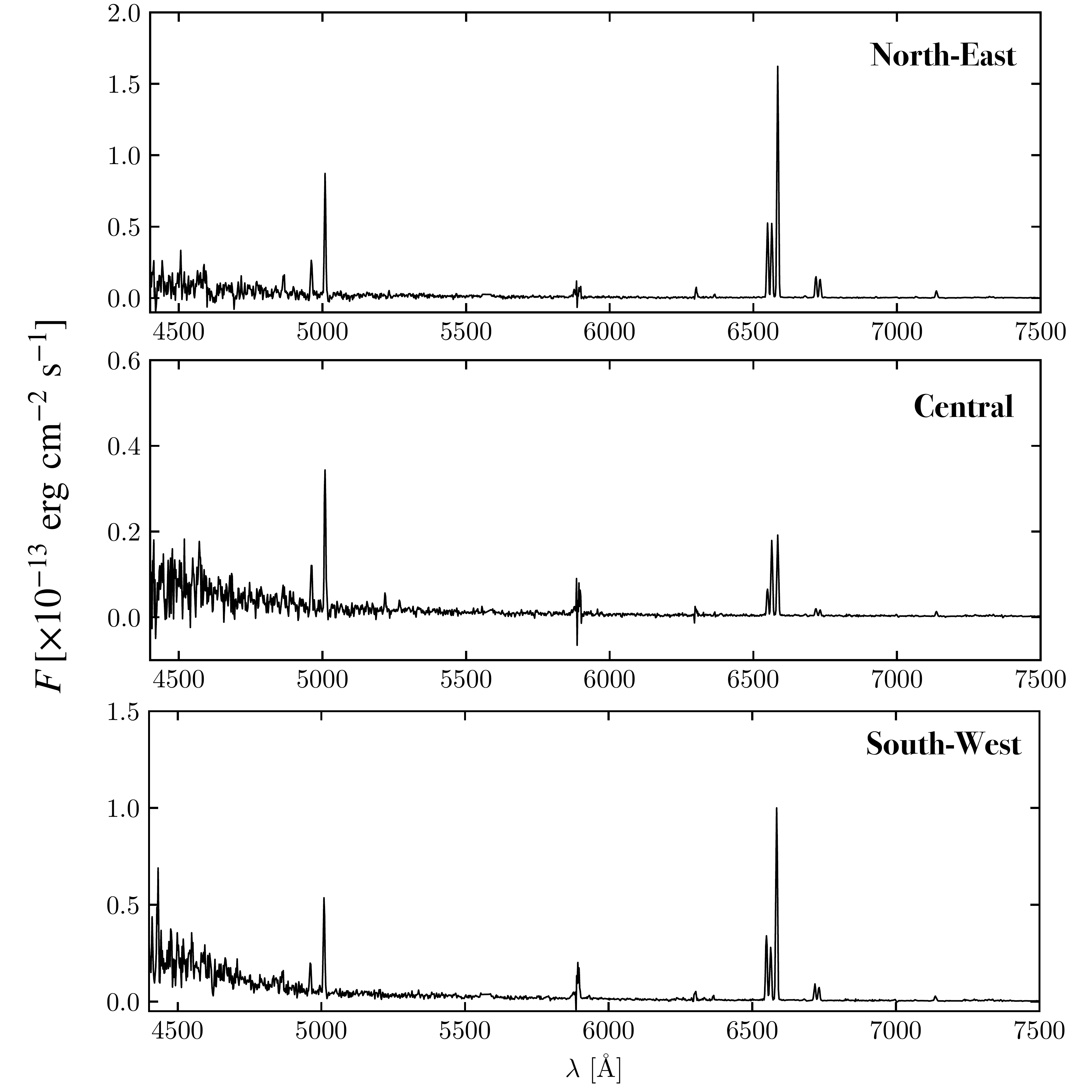}
\caption{GTC OSIRIS deredenned spectra of IPHASX J191104.8+060845. The
  upper, middle and bottom panels show the spectra of the North-East,
  central and South-west region of the main ellipsoidal structure.}
\label{fig:Sab2_opt}
\end{center}
\end{figure*}

In order to peer further into the structure of J191104, we have
extracted surface brightness profiles from the three nebular
images. We have selected PAs of $+47^{\circ}$, $-10^{\circ}$ and
$-55^{\circ}$ to extract 1D profiles from the three narrow-band filter
images. These three PAs trace the surface brightness distribution of
the central ring, the direction of the A--A$'$ knots and that of the
C--C$'$ knots, respectively. All profiles are presented in
Figure~\ref{fig:Sab2_profiles}.  This figure shows that the [N\,{\sc
    ii}] emission encloses that of H$\alpha$ and [O\,{\sc iii}]. The
elliptical structure is illustrated by the leftmost panel of
Figure~\ref{fig:Sab2_profiles}. The profiles of the three emission
lines show that the south-west region of the ellipse is broader than
the north-east region with FWHM of $\sim5''$ for the [N\,{\sc ii}]
emission profile.

The middle panel of Figure~\ref{fig:Sab2_profiles} shows that the
A--A$'$ pair of knots are broad and comparable to that of the main
elliptical shell profile. They exhibit a FWHM of $\sim4-5''$. This
pair of knots is also marginally detected in the H$\alpha$ emission
profile (see Fig.~\ref{NOT}). Figure~\ref{fig:Sab2_profiles} also
shows that the A--A$'$ pair of knots are as bright as the elliptical
main shell in J191104. On the other hand, the rightmost panel of
Figure~2 shows that the C--C$'$ pair is dimmer. Finally, we note that
the middle and right panels of Figure~\ref{fig:Sab2_profiles} also
show that the H$\alpha$ exhibit a center-filled morphology filling the
inner [N\,{\sc ii}] shell.

\subsection{Chemistry}

We extracted three spectra from the GTC spectrum, two corresponding to
the bright rims of the central ring denoted as North-East (NE) and
South-West (SW) and another from the central region of the main cavity
in J191104. These are shown in Figure~\ref{fig:Sab2_opt}. According to
the images presented in Figure~\ref{NOT}, the spectra are dominated by
the [N\,{\sc ii}] line at 6584~\AA. The contribution from other
emission lines is smaller (see Fig.~\ref{fig:Sab2_opt}). All detected
emission lines in the spectra are listed in Table~\ref{tab:Fluxes}.

Due to the similarity between the North-East and South-West spectra,
we only present the analysis of the former. The spectrum is the one
used to analyse the physical and chemical properties of J191104. The
analysis tool {\sc anneb} \citep{Olguin2011}, based on the {\it
  Nebular}/{\sc iraf} package \citep{Shaw1995}, was used to derive the
parameters of the nebula (this includes using the atomic data from
{\sc iraf} libraries). The logarithmic extinction ¡ derived from
H$\alpha$/H$\beta$ Balmer decrement is found to be particularly high,
with $c$(H$\beta$) = 3.36$\pm$0.16. This corresponds to a reddening
$E(B-V)=2.77\pm$0.13 in concordance with the extinction values in this
line of sight given in the Galactic DUST Reddening and
Extinction\footnote{\url{https://irsa.ipac.caltech.edu/applications/DUST/}}
at the NASA/IPAC Infrared Science Archive.  The dereddening of the
emission lines was performed using the extinction curve from
\citet{Fitzpatrick2007} for $R_\mathrm{V}$=3.1 Details of the line
fluxes are listed in Table~\ref{tab:Fluxes}.

The logarithmic flux ratios shown in Table \ref{FR} are used to
position our target in updated diagnostic diagrams by \citet{Frew2010}
and \citet{Sabin2013}, and confirm the PN nature of J191104.8, though
morphology alone in these new deeper narrow-band images are already
highly indicative. The source is characterised by a low electronic
temperature, $T_\mathrm{e}$([N\,{\sc ii}])= 8000$\pm$1800~K, and a low
density of $N_\mathrm{e}=260 \pm 20$~cm$^{-3}$ as estimated from the
[S\,{\sc ii}] $\lambda\lambda$6717,6731 doublet. We note that our
density estimate is at the edge of the physical lower limit derived by
this method \citep{Osterbrock}. J191104 has too few lines to perform a
detailed abundance analysis, but we were able to obtain some mean
ionic abundances for the North-East spectrum. These are listed in
Table~\ref{tab:ionic}. Due to the low ionization state of the nebula
we do not expect species with very high ionization degree. Based on
the ionic abundances results and assuming that these can be
approximated to the total abundances (e.g. N$^{+}$/O$^{+}$\,$\sim$
N/O) it appears that J191104.8 is rich in He, O and N.

Based on the spectroscopic analysis we are able to assert the PN
nature of J191104. Unfortunately, due to the low emission we cannot
derive the total abundances, but we can use the spectra and ionic
abundances (using H$^+$) to classify this PN. To do so, we consider
the logarithmic flux ratios shown in Table \ref{FR} to position our
target in diagnostic diagrams. J191104 falls in the region of Type I
PNe \citep[][]{Peimbert1978,Peimbert1980}. These objects are
characterized by their enrichment in nitrogen and helium, with
abundance ratios He/H$\geq$0.14 and log(N/O)$\geq-0.3$. It is believed
that the excess in N and He abundance is the result of the evolution
of progenitors more massive than 2.4~M$_{\odot}$, which implies that
it was recently formed from a medium that is presumably richer in
metals and helium \citep{Peimbert1987}. Furthermore, these nebulae
have many filamentary structures and usually show a bipolar
morphology. We found that J191104 has log(N/O) = 0.74$\pm$0.14 and
according to Peimber's criterion, this PN agrees with a Type I PN.

\begin{table}
\begin{threeparttable}
\begin{center}
\caption{Emission lines detected in the GTC OSIRIS North-East spectrum of J191104.8+060845$^\dagger$.}
\begin{tabular}{cccc}
\hline
Ion & $\lambda_0$ & $F$  & $F^{\star}$  \\
\hline
H$\beta^\dagger$        &  4861  & 100  $\pm$ 24 & 100 $\pm$ 24 \\
\,[O\,{\sc iii}]&  4959 & 208 $\pm$ 38 & 173 $\pm$ 32\\
\,[O\,{\sc iii}]&  5007 & 735 $\pm$ 126 & 559 $\pm$ 97 \\
\,[N\,{\sc ii}] &  5755 & 26 $\pm$ 30 & 6 $\pm$ 7 \\
He\,{\sc i}     & 5876  & 157 $\pm$ 37 & 28 $\pm$ 8 \\
\,[O\,{\sc i}]  & 6300  & 333 $\pm$ 60 & 34 $\pm$ 9 \\
\,[O\,{\sc i}]  & 6363  & 110 $\pm$ 24 & 10 $\pm$ 3 \\
\,[N\,{\sc ii}] & 6548  & 3,833 $\pm$ 655 & 287 $\pm$ 85\\
\,H$\alpha$     & 6563  & 3,883 $\pm$ 663 & 286$\pm$ 55\\
\,[N\,{\sc ii}] & 6583  & 12,300 $\pm$ 2,100 & 885 $\pm$ 263 \\
He\,{\sc i}     & 6678  & 107 $\pm$ 28 & 7 $\pm$ 3\\
\,[S\,{\sc ii}] & 6716  & 1,307 $\pm$ 224 & 81 $\pm$ 25 \\
\,[S\,{\sc ii}] & 6731  & 1,221 $\pm$ 209 & 74 $\pm$ 23\\
He\,{\sc i}     & 7065  & 102 $\pm$ 24 & 4 $\pm$ 2 \\
\,[Ar\,{\sc iii}]& 7136 &  652 $\pm$ 112 & 26 $\pm$ 9 \\
\,[O\,{\sc ii}] & 7320  & 136 $\pm$ 26 & 5 $\pm$ 2  \\
\,[O\,{\sc ii}] & 7330  & 81 $\pm$ 21 & 3 $\pm$ 1 \\
\hline
\end{tabular}
\label{tab:Fluxes}
\begin{tablenotes}
      \small
      \item $^{\star}$Dereddened values.
      \item $^\dagger$ All values are normalized to $F$(H$\beta$)=100 with $F_\mathrm{H\beta}=(4.39\pm0.75)\times10^{-17}$
      \end{tablenotes}
\end{center}
  \end{threeparttable}
\end{table}

\begin{table}
\begin{center}
\caption{Flux ratios for the diagnostic diagrams }
\begin{tabular}{|l|c|c|}
\hline
Species   & Value & Error \\
\hline
log(H$\alpha$/[N\,{\sc ii}])  & $-$0.6 & 0.1\\
log(H$\alpha$/[S\,{\sc ii}])  & 0.3    &  0.1  \\
log([O\,{\sc iii}]/H$\alpha$) & 0.4    & 0.1 \\
\,[S\,{\sc ii}] 6716/6731     & 1.2   & 0.3 \\
\hline
\end{tabular}
\label{FR}
\end{center}
\end{table}

\begin{table}
\begin{center}
\caption{Mean ionic abundances obtained from the GTC OSIRIS North-East spectrum of J191104.8+060845.}
\begin{threeparttable}
\begin{tabular}{|l|c|c|l}
\hline
Ion & $N_\mathrm{ion}/N_\mathrm{H+}$ & error (\%)$^\star$ &  $\lambda$ (\AA)$\dagger$ \\
\hline
He$^{+}$ &  0.2                & 17.0     & 5876,6678,7065\\
N$^{+}$ &  $2.0\times10^{-4}$  & 15.2     & 5755,6548,8583\\ 
S$^{+}$ &  $2.0\times10^{-6}$  & 10.0     & 6716,6731 \\ 
O$^{2+}$ & $2.0\times10^{-4}$  & 14.0     &  4959,5007\\ 
O$^{+}$ &  $3.0\times10^{-5}$  & 109.0    & 7320,7330\\
Ar$^{2+}$ & $5.00\times10^{-6}$  & 22.5   & 7136\\ 
\hline
\end{tabular}
\begin{tablenotes}
     \small
     \item $^\star$The uncertainties in the ionic abundances are the
       result of propagating the line fluxes errors and those of
       electron temperature into the estimations.
    \item $^\dagger$Wavelength of the emission lines used to compute
      the ionic abundance.
\end{tablenotes}
\end{threeparttable}
\label{tab:ionic}
\end{center}
\end{table}

\subsection{Kinematics}

\begin{figure}
\begin{center}
\includegraphics[angle=0,width=\linewidth]{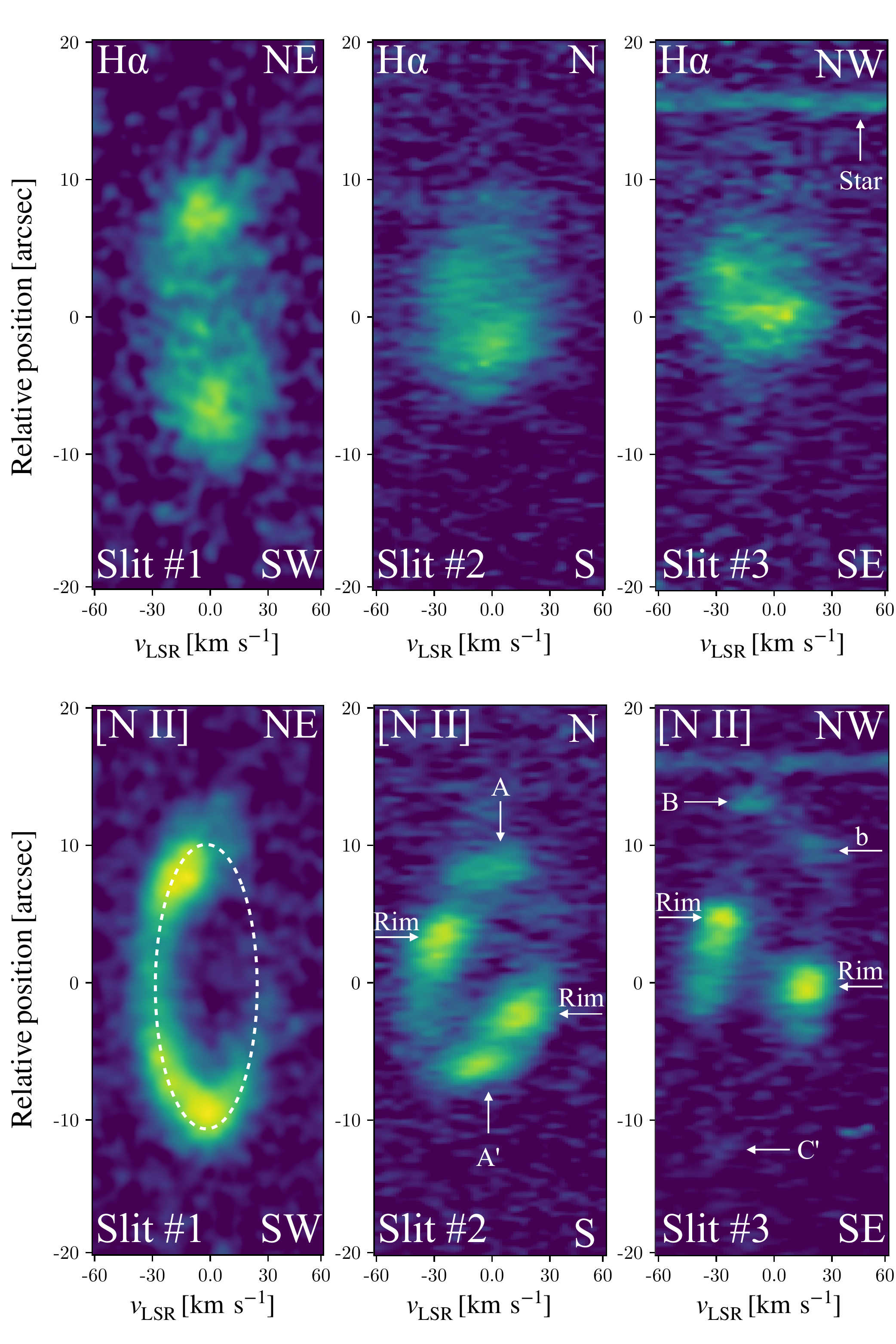}
\caption{Echelle spectra obtained with the MES at SPM. From left to
  right, Slit \#1, \#2 and \#3. The upper panels show the H$\alpha$
  emission line and the bottom their corresponding [N\,{\sc ii}]
  emission. Note the contribution from a background star in the
  Slit\,\#3 spectra.}
\label{fig:Sab2_echelle}
\end{center}
\end{figure}

The position-velocity (PV) diagrams obtained from our SPM-MES
observations are shown in Figure~\ref{fig:Sab2_echelle}. The top
panels present the H$\alpha$ emission line extracted from the three
slit positions shown in Figure~\ref{NOT}, while the bottom panels of
this figure present the corresponding [N\,{\sc ii}] emission line
profiles. We first notice that the [N\,{\sc ii}] profiles clearly
trace well-defined structures, whilst the H$\alpha$ profiles are
fuzzy, thus in the following we describe the kinematics of J191104
only taking into account the [N\,{\sc ii}] line profiles.

The Slit~1, which was obtained with a PA=47$^{\circ}$, unveils the
kinematics of the main cavity of J191104. This profile discloses the
kinematic signature of an expanding ellipsoidal structure (Fig.~4
bottom left panel). The line profile lacks emission in the red shifted
part of the spectrum. The expansion velocity obtained by fitting an
ellipse to the line profile (see Fig.~\ref{fig:Sab2_echelle} bottom
left panel) is 26~km~s$^{-1}$.

Slits~2 and 3 trace the velocity components of the central ring-like
structure and are labeled accordingly in
Figure~\ref{fig:Sab2_echelle}. They also exhibit some structures,
which extend NW--SE according to the nebular images presented in
Figure~\ref{NOT}. The Slit~2 (PA=$-10^{\circ}$), which covers the
positions of the A--A' pair of knots, shows that the brightest region
of these features have velocities close to zero, although they extend
from $\sim20$~km~s$^{-1}$ to $\sim-$27~km~s$^{-1}$ with respect to the
systemic velocity of the main shell.

Slit~3 (PA=$-43^{\circ}$) covers the positions of the features labeled
as B, b and C$'$. The bottom right panel of
Figure~\ref{fig:Sab2_echelle} suggests that the knots B and b are part
of a lobe that expands towards the NW protruding from the inner
ring. Their velocity difference is $\sim$60~km~s$^{-1}$. The C$'$ knot
is marginally detected in the panel at a velocity of
$-28$~km~s$^{-1}$. The expansion velocity of the ring obtained from
Slits~2 and 3 are 23.5~km~s$^{-1}$ and 23~km~s$^{-1}$, respectively,
which are consistent with that estimated from Slit~1.

\begin{figure}
\begin{center}
  \includegraphics[angle=0,width=\linewidth]{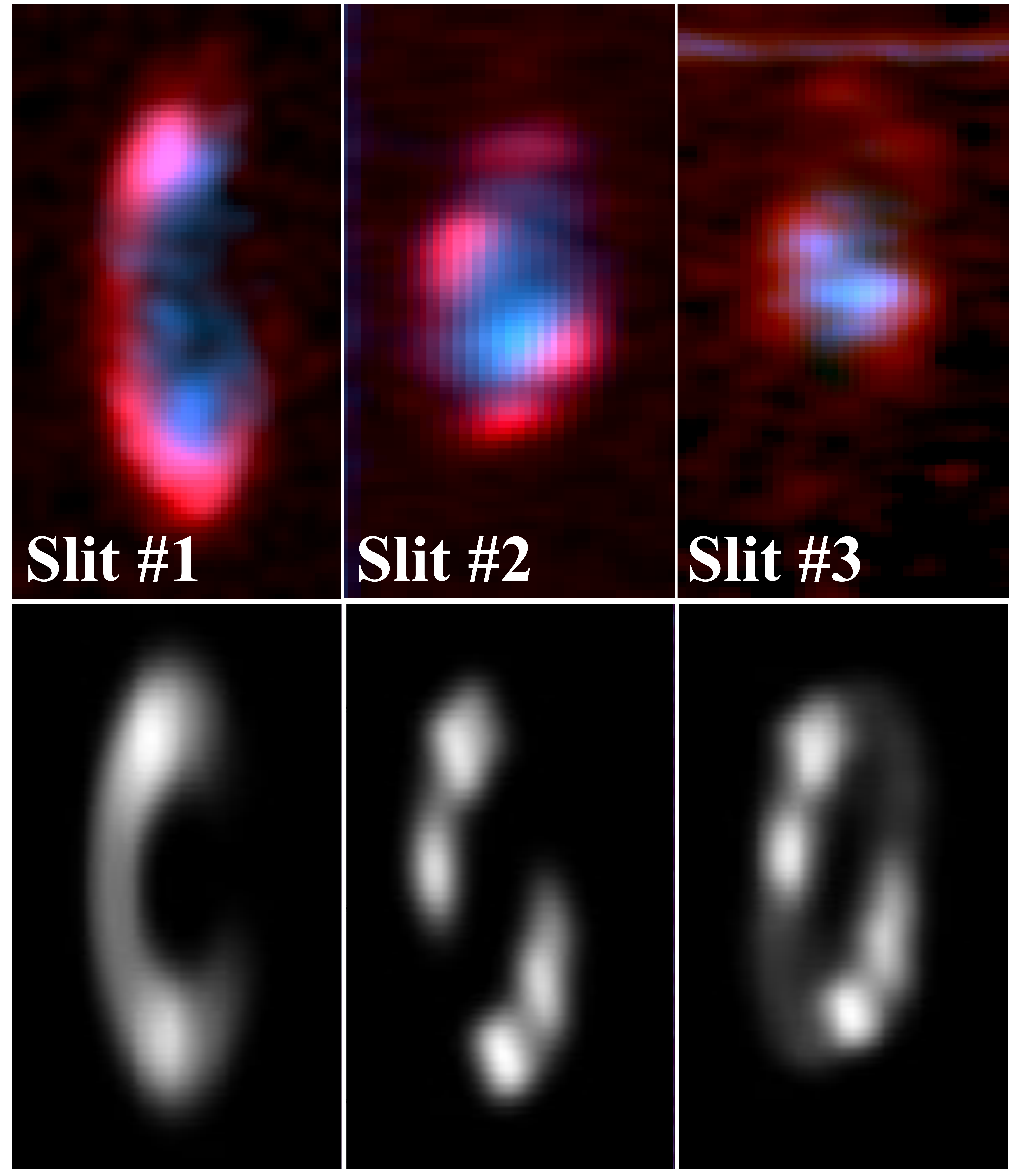}
\caption{Colour-composite images of the PV shown in Fig.~4. In the top
  panel, red and blue correspond to the emission from [N\,{\sc ii}]
  and H$\alpha$, respectively. The bottom panels show the synthetic PV
  images from our best {\sc shape} morpho-kinematic model obtained in
  section 4.2}
\label{fig:Sab2_echelle_RGB}
\end{center}
\end{figure}

In Figure~\ref{fig:Sab2_echelle_RGB} we present colour-composite
images combining the spectra presented in
Figure~\ref{fig:Sab2_echelle}. This figure shows that although the
structures are not traced in detail by the H$\alpha$ profiles, the
H$\alpha$ emission is contained inside the [N\,{\sc ii}] emission. Gas
that emits in this line has thus lower expansion velocities than that
of the [N\,{\sc ii}]-emitting gas.

\section{Discussion}

The images and spectra presented in the previous sections show that
J191104 is a PN with low-ionization structures. Its emission is
dominated by the [N\,{\sc ii}] line and the PN shows a small
contribution from the [O\,{\sc iii}] line. In order to unveil its
origin we have performed several analysis to try to assess the
characteristics of this PN and its central star.

\subsection{Physical properties and structure of J191104}

To determine the physical size of J191104 we need to estimate the
distance to this PN. It is impossible to determine the distance using
the central star because, as we show in the following section,
identifying the progenitor star is a difficult task. Thus, we need to
use indirect methods such as the one proposed by
\citet{Frew2016}. These authors developed a statistical method to
calculate distances to PNe using the H$\alpha$ surface brightness
($S_\mathrm{H\alpha}$). This quantity can be defined as
\begin{equation}
S_\mathrm{{H\alpha}}= \frac{F_\mathrm{H\alpha}}{A}, 
\end{equation}
\noindent 
Where $A$ represents the angular area of the object in steradians (sr)
and $F_\mathrm{H\alpha}$ is the flux of the object in the H$\alpha$
emission line in units of erg~cm$^{-2}$~s$^{-1}$. Thus, we need to
estimate the total H$\alpha$ luminosity and surface brightness of
J191104. For this we used our GTC OSIRIS spectra of the central region
of J191104. The H$\alpha$ flux was scaled to an ellipsoidal region
with semi-axis of 11$''\times$8$''$, which encompasses the main cavity
in J191104. We found an H$\alpha$ surface brightness of
$S_\mathrm{H\alpha}=2.84\times10^{-5}$~erg~cm$^{-2}$~s$^{-1}$~sr$^{-1}$.

\begin{figure}
\begin{center}
  \includegraphics[angle=0,width=\linewidth]{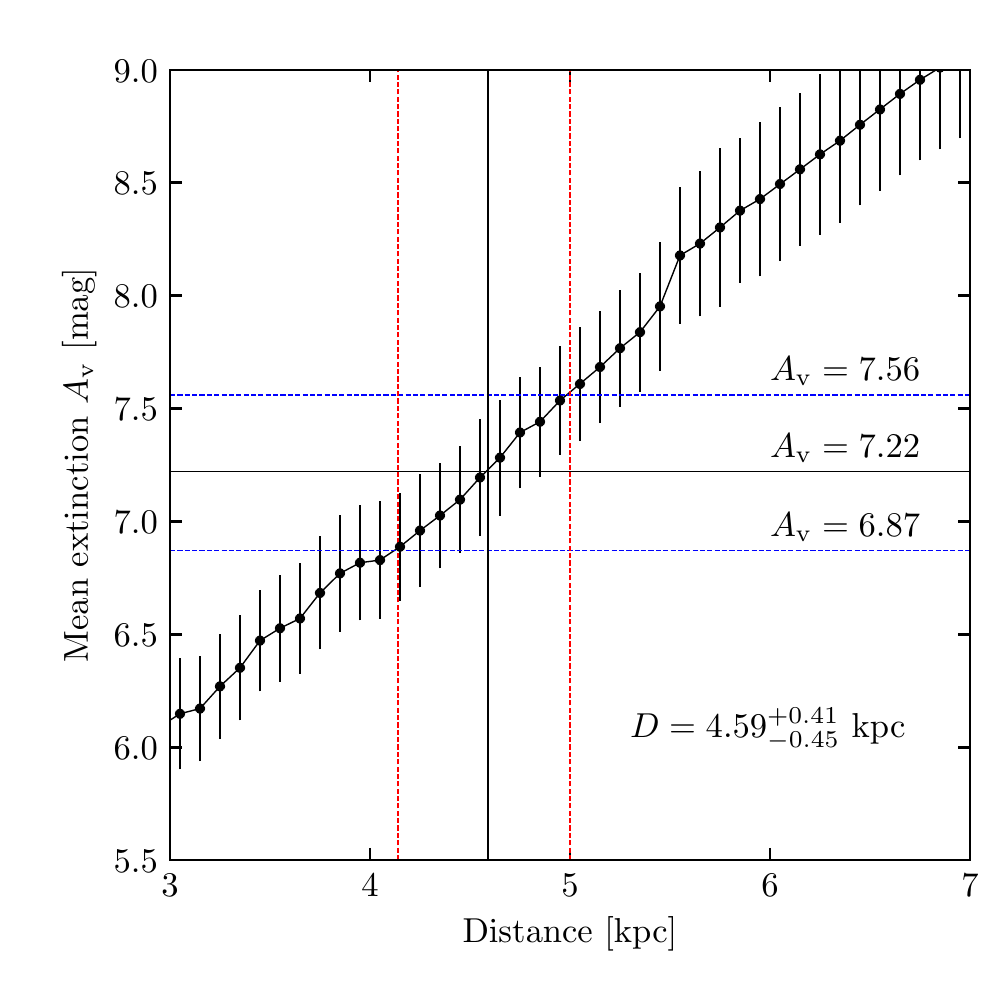}
\caption{Estimated distance as a function of mean extinction
  $A_\mathrm{v}$ following the method described by \citet{Sale2014}.}
\label{fig:dist_Av}
\end{center}
\end{figure}

\begin{figure*}
\begin{center}
\includegraphics[angle=0,scale=0.8]{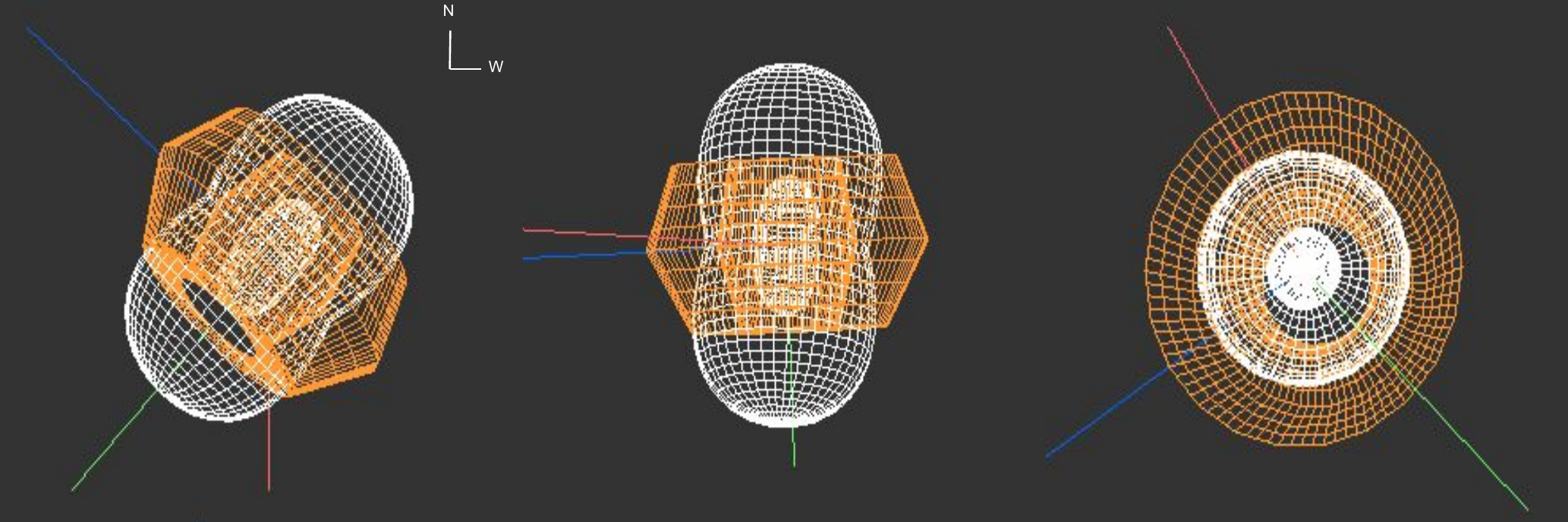}
\caption{Morpho-kinematic {\sc shape} model of
  IPHASXJ\,191104.8+060845 based on PV diagrams presented in
  Figure~4.The left panel corresponds to the nebula projected onto the
  plane of the sky, in the central panel we have a side view and in
  the right panel a view through the lobe.}
\label{fig:shape_model}
\end{center}
\end{figure*}

\citet{Frew2016} presented different relationships between
$S_\mathrm{H\alpha}$ and the size ($r$) according to the
characteristics of the PN. One of the criteria to determine the type
of nebula is the flux ratio between the [N\,{\sc ii}]~6584~\AA\, and
H$\alpha$ emission lines. If F([N\,{\sc
    ii}]6584\AA)/F(H$\alpha$)$\geq$1 we are in the optically thick
regime for PNe. This is the case for our PN where we find a ratio of
3.3. The corresponding $S_\mathrm{H\alpha}-r$ relation for this type
of nebula is:
\begin{equation}
\mathrm{log}\,S_\mathrm{H\alpha } = -(3.32\pm0.12)\mathrm{log}\,r - (4.97\pm0.08).
\end{equation}
\noindent Using our estimate of the $S_\mathrm{H\alpha}$ we obtained $r=0.75\pm0.04$~pc. Then, the distance can be calculated as
\begin{equation}
D \approx 206265 \left(\frac{r}{\mathrm{pc}}\right) \left(\frac{\theta}{''}\right)^{-1} \, \mathrm{pc},
\end{equation}
\noindent where $\theta$ is in this case the largest extension of
J191104, using scale plate we find $\theta=30''$. By adopting this
value for $\theta$ we are assuming the somewhat evident fact that the
H$\alpha$ emission from the central region of J191104 dominates the
total H$\alpha$ flux. Then, the distance to J191104 is
$D=5.1\pm0.3$~kpc.

Another way of determining the distance is calculating the extinction
towards stars in the vicinity of J191104 following the method
described by \citet{Sale2014}. This is based on the relationship
between the extinction of a source and that of field's stars in the
IPHAS sky with known distances located in the source line of
sight. The extinction map is illustrated in Figure~\ref{fig:dist_Av}
and we used A$_{v}$= 2.15$\times$ $c$(H$\beta$) for coherence with the
maps generation method. The resulting distance is
$D_{2}=4.6^{+0.4}_{-0.5}$~kpc which is consistent with that derived
following above.

Using the two independent distance estimates we can calculate an
averaged distance of $D = 4.9\pm0.6$~kpc and we can now proceed to
calculate a dynamical age for the morphological components in
J191104. The inner ring has an extension of 0.26~pc and using our MES
observations we found that the main cavity is expanding with a
velocity of 23$\pm$3~km~s$^{-1}$. Thus a kinematical age, $\tau$ can
be obtained as:
\begin{equation}
\tau \approx 978000 \left( \frac{r}{\mathrm{pc}} \right) \left( \frac{v}{\mathrm{km/s}} \right)^{-1}.
\end{equation}
\noindent We find that J191104 has a kinematic age $\tau$=11,000$\pm$1,500~yr.

\subsection{Morpho-kinematics Modelling}

In order to interpret the MES observations and to unveil the kinematic
structure of J191104 we have used {\sc shape}
\citep[][]{Steffen2011}. {\sc shape} is used to reconstruct and model
the morpho-kinematic signatures of astrophysical objects. Using this
software we can build a morpho-kinematic model that reproduces the PV
diagrams in Figure~\ref{fig:Sab2_echelle}. Our best model was achieved
by a main barrel-like structure with an inclination of 42$^{\circ}$,
which corresponds to the central part of the nebula showing the
highest emission. It is also necessary to add a hollow bipolar
structure with a similar inclination. In Figure~\ref{fig:shape_model}
we present the visualization of our {\sc shape} model with different
perspectives. Synthetic spectra obtained from similar positions as
those shown in Figure~\ref{NOT} are compared to the MES observations
in Figure~\ref{fig:Sab2_echelle_RGB}. We note that the observations
obtained from slit 3 might suggest that the NW lobe has sub-structures
(see Fig.~\ref{fig:Sab2_echelle} and \ref{fig:Sab2_echelle_RGB}).
This might be the same case for the SE lobe, but self absorption due
to dust in this PN (see next section) hampers a stronger
statement. Furthermore, a better representation of the observations is
severely hampered by the fact that the faint structures are below
detection limits of MES.

\subsection{Searching for a progenitor}

\begin{figure*}
\begin{center}
\includegraphics[angle=0,width=0.45\linewidth]{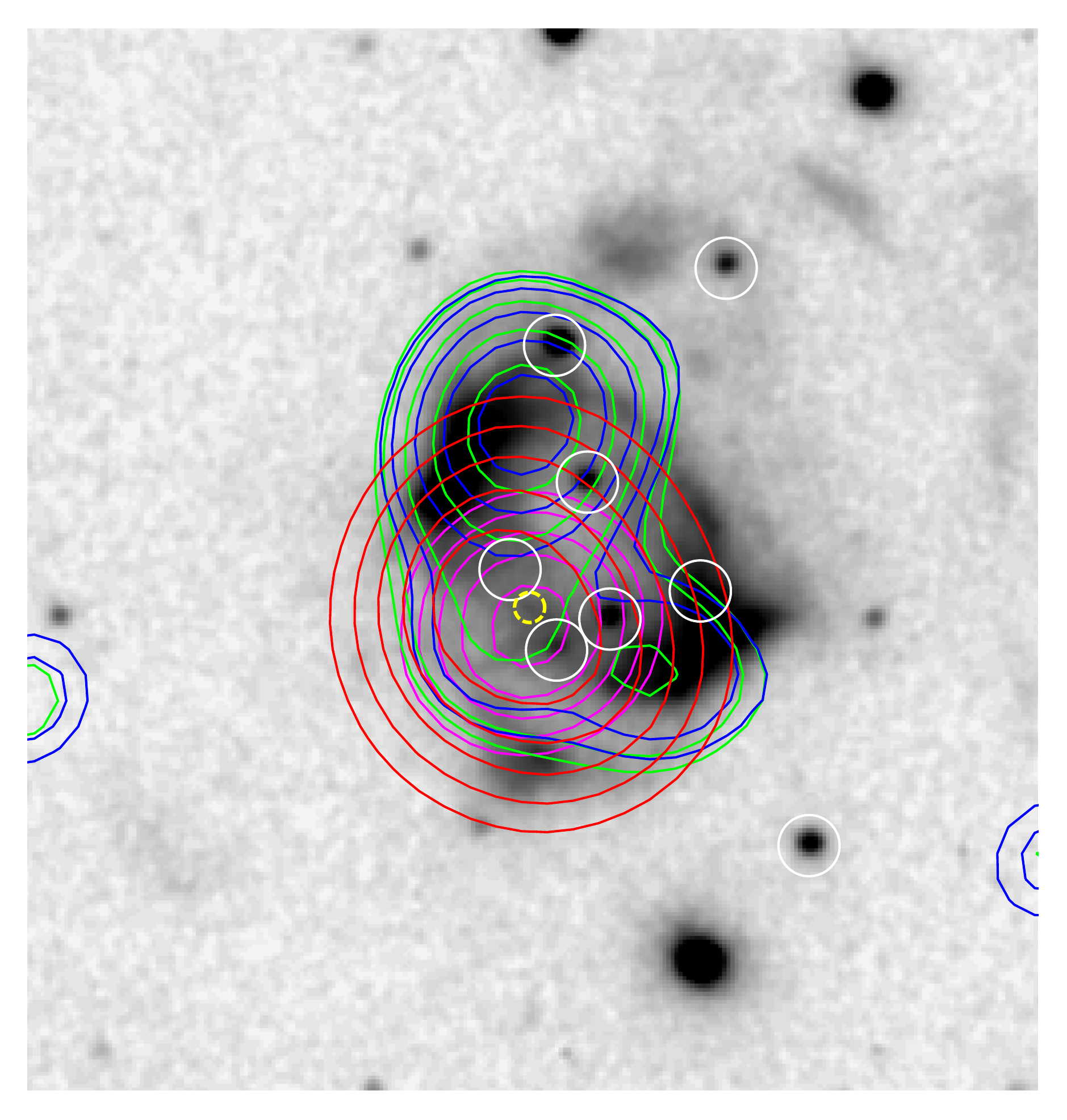}~
\includegraphics[angle=0,width=0.45\linewidth]{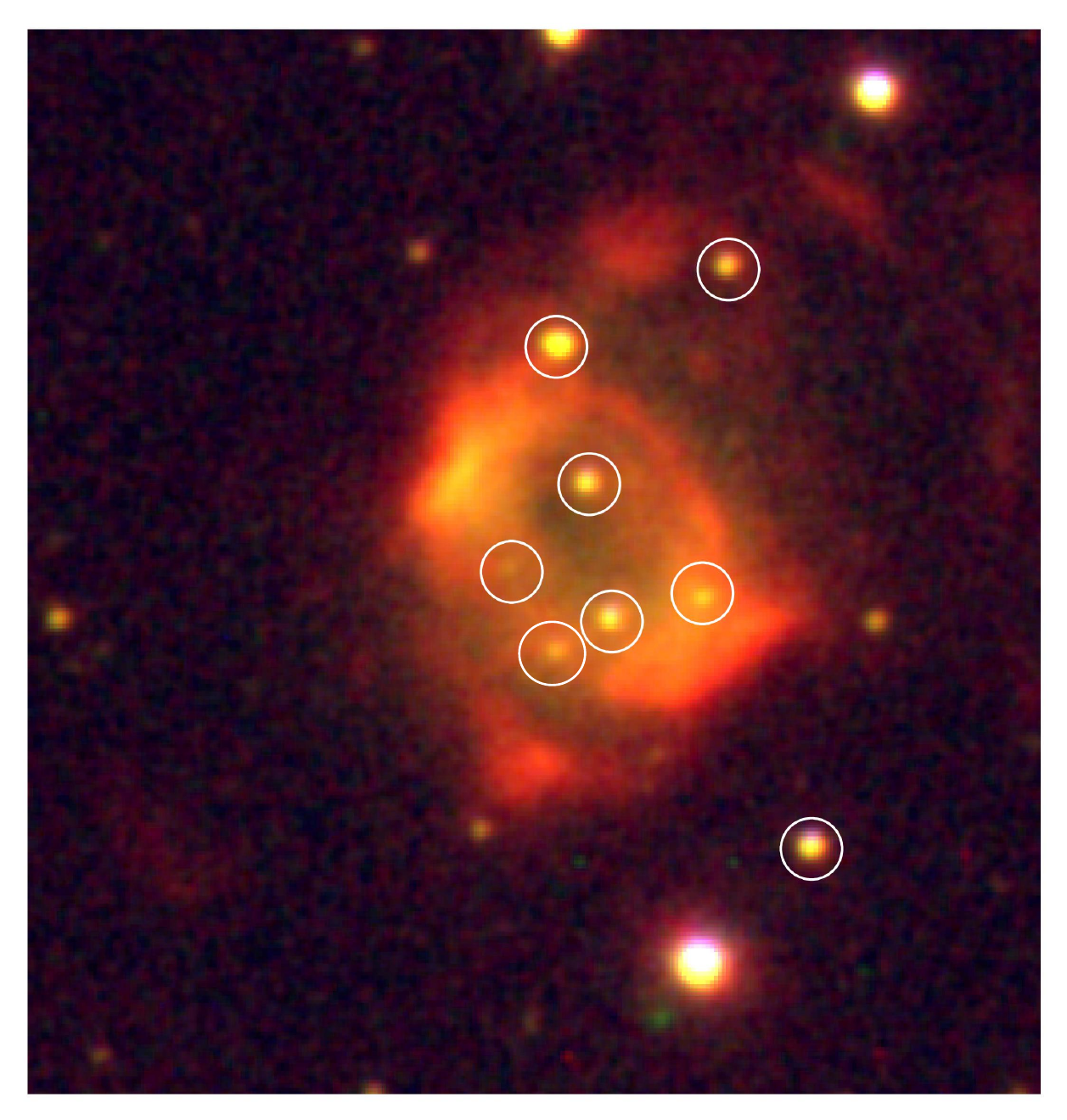}
\caption{Comparison between IR and optical emission from IPHASX
  J\,191104.8+060845. Left: Grey-scale [N\,{\sc ii}] image. The
  magenta, red, green and blue contours correspond to the near-IR
  emission detected by {\it WISE} bands W4~22~$\mu$m, W3~12~$\mu$m,
  W2~4.6~$\mu$m, and W1~3.4~$\mu$m, respectively. Right:
  Colour-composite optical image similar to Figure~1. The white
  circles show the positions of stars detected in the optical and {\it
    2MASS} images. North is up, East to the left.}
\label{fig:compare}
\end{center}
\end{figure*}

Identifying the progenitor star of J191104 is not a trivial task. One
might have expected that as a first approximation the progenitor star
should be projected (or contained) within the inner cavity of
J191104. There are two apparent stars that come close to this
condition (see Fig.~\ref{NOT}). However, these two sources are located
close to the rim of the inner cavity making them unlikely to be the
progenitor star.

As a first step to try to idenfity the progenitor star of J191104 we
searched the UV-Excess survey
\citep[UVEX;][]{Groot2009}\footnote{\url{https://www.astro.ru.nl/uvex/Science.html}}. This
is a complementary survey to the IPHAS which maps the Northern
Galactic Plane ($b < 5^{\circ}$) with the 2.5~m INT but with the $U$,
$g$, $r$, and He\,{\sc i}~5875 filters. UVEX looks for hot, blue
objects, with relatively low luminosity. However, looking at the UVEX
survey data, we could not find any noticeable central star. Also
PanSTARRS \citep[][]{Chambers2016} $g$ band shows no signs of
detection either.Based on the weakest field star in PanSTARSS, we
establish that the central star will have a magnitude dimmer than
21. This tells us that the central star must be an evolved low
luminosity object that might also be obscured by dust in the nebula.

Furthermore, we also searched the IR catalogues at
IRSA\footnote{\url{https://irsa.ipac.caltech.edu/frontpage/}} for
obscured candidates. The {\it 2MASS} images of the vicinity of J191104
revealed the presence of point-like sources which are coincident with
those detected in optical images (see Fig.~\ref{fig:compare}). A
search in the {\it WISE} data of the vicinity of J191104 show no
apparent correlation with point sources but extended mid-IR
emission. The {\it WISE} W3 at 12~$\mu$m and W4 at 22~$\mu$m images
exhibit a single maximum associated with the inner region of the main
cavity. If one takes into account the morpho-kinematic model presented
in the previous section (see the innermost structure in the right
panel of \ref{fig:shape_model}), we can suggest a position for the
central star, and thus assume that it is placed close to the geometric
center of the nebula (i.e. slightly off centered). This is also
illustrated with a yellow dashed-line circle in the left panel of
Figure~\ref{fig:compare}. This position seems to be coincident with
the {\it WISE} W3 and W4 bands maxima which might point at the
location of the progenitor star of J191104, although the resolution is
not sufficient to strongly constrain this suggestion.

On the other hand, the {\it WISE} W1 at 3.4~$\mu$m and W2 at
4.6~$\mu$m bands exhibit morphologies very similar as those traced by
the [N\,{\sc ii}] NOT image (Figure~\ref{fig:compare}). Two near-IR
maxima are associated to the NE and SW edges of the inner cavity of
J191104. For comparison we show in the left panel of
Figure~\ref{fig:compare} the [N\,{\sc ii}] emission compared with
contours obtained from the four {\it WISE}
bands. Figure~\ref{fig:compare} points at the presence of dust
associated to the main cavity structure in J191104 which might be
responsible for extinguishing the brightness of the SE lobe.

Here we have shown that J191104 is an old, evolved , heavily reddened
and extincted PNe. The present work suggests that dust has survived
the evolution of the CSPN for 11,000~yr. This means that the
barrel-like structure might have been dense enough to shield dust from
the UV flux. Another possibility is that dust has recently formed due
to the reduction of the ionizing photon flux from the progenitor
star. This is an interesting idea worth pursuing with the
characterization of IPHAS objects. The very late stages of evolution
of PNe have not been studied in detail. For example, what are the
observable properties of PNe in which the power of the stellar wind
has diminished \citep[e.g.,][]{GS2006}.

\section{Conclusion}

We have presented a multi-frequency study of the nebula IPHASX
J191104.8+060845 (J191104) originally detected in the IPHAS. We have
used images and spectra to demonstrate that this is in fact a bipolar,
heavily-extincted, old PN. Our findings can be summarized as:
\begin{itemize}

\item Using GTC OSIRIS spectra we found that the log(N/O) ratio is
  $0.74\pm0.14$ and He/H $\gg$ 0.20. Thus it can be classified as a
  Peimbert Type I, which means that J191104 has been formed by the
  evolution of a massive progenitor star very likely more massive than
  2.4~M$_{\odot}$. The electron density and temperature are found to
  be $N_\mathrm{e}=200\pm 20$~cm$^{-3}$ and $T_\mathrm{e}=8000 \pm
  1800$~K.

\item Our NOT images unveiled the presence of a bright central
  elliptical structure surrounded by detached patches of emission
  mostly detected in [N\,{\sc ii}] and marginally in the H$\alpha$
  image. Some of the clumps seem to be paired and are located towards
  the NW and SE from the main inner cavity. Faint extended bipolar
  lobes are also revealed for the first time.

\item Using two independent methods we estimate that the distance to
  J191104 is $D\approx4.9$~kpc and with the help of the MES
  observations we calculated a kinematical age of 11,000~yr.

\item The kinematic data obtained from our MES observations help us
  reconstruct, as a first approximation, the structure and kinematics
  of J191104. For this, we used the interactive 3D morpho-kinematic
  application {\sc shape}. Our best model of J191104 suggests that the
  PN is composed by a main barrel-like dense structure with a hollow,
  bipolar structure. The morphology of the structures towards the NW
  of the main cavity suggests that several lobes might exist in
  J191104. The high extinction revealed by the optical and IR images
  might hamper the detection of the SE lobes

\item We did not identify a progenitor star using the UVEX catalogue,
  but the analysis of IR observations agree with the suggested
  position for the progenitor star (assuming that it should be located
  at the geometric center of J191104). Furthermore, the IR
  observations suggest that the dust is spatially correlated with the
  main cavity in J191104. Either dust has survived the evolution of
  the central star for the past 11,000~yr or it has recently formed
  due to the reduction of the UV flux of the progenitor star.
\end{itemize}

IPHAS is giving us the unique possibility of studying old and/or
extinguished objects in the North Galactic plane. Although much
attention has been paid to aspherical, young PNe with the {\it Hubble
  Space Telescope} \citep[see][and references
  therein]{Hsia2014,Sahai2011} this is the missing link to study the
fate of the circumstellar matter around solar-like stars.  We will be
able to study phases in which the stellar wind dims and the UV flux
practically disappears leaving the ionized gas to evolve under its own
inertia. These evolved PNe will also allow us to assess whether dust
survives the evolution of the PNe or if dust is being formed due to
the reduction of the UV flux from the CSPN.

\section*{Acknowledgements}

The authors are thankful to the referee for a prompt report that
improved the presentation of the present paper.  J.B.J.-G. and J.A.T.
are funded by UNAM DGAPA PAPIIT projects IA100318 and
IA100720. J.B.J.-G. and L.S. are funded by UNAM-PAPIIT grant IN101819.
SZ works under the collaboration agreement "UNAM-TecNM
43310-3020-30-IX-15".  GR-L acknowledges support from CONACyT and
PRODEP (México).  MAG acknowledges support from the Spanish Government
Ministerio de Ciencia, Innovaci\'on y Universidades through grant
PGC2018-102184-B-I00.  QAP thanks the Hong Kong Research Grants
Council for GRF research support under grants 17326116 and
17300417. We thank the daytime and night support staff at the OAN-SPM
for facilitating and helping obtain our observations.  This paper
makes use of data obtained as part of the INT Photometric H$\alpha$
Survey of the Northern Galactic Plane (IPHAS) carried out at the Isaac
Newton Telescope (INT). The INT is operated on the island of La Palma
by the Isaac Newton Group in the Spanish Observatorio del Roque de los
Muchachos of the Instituto de Astrof\'isica de Canarias. All IPHAS
data are processed by the Cambridge Astronomical Survey Unit, at the
Institute of Astronomy in Cambridge.  Based on observations collected
at the Observatorio Astron\'omico Nacional at San Pedro M\'artir,
B.C., Mexico. Based on observations made with the Nordic Optical
Telescope, operated by the Nordic Optical Telescope Scientific
Association at the Observatorio del Roque de los Muchachos, La Palma,
Spain, of the Instituto de Astrofisica de Canarias.  Based on
observations made with the Gran Telescopio Canarias (GTC), installed
at the Spanish Observatorio del Roque de los Muchachos of the
Instituto de Astrofísica de Canarias, in the island of La Palma. The
data presented here were obtained in part with ALFOSC, which is
provided by the Instituto de Astrof\'{i}sica de Andaluc\'{i}a (IAA)
under a joint agreement with the University of Copenhagen and NOTSA.

\section*{Data availability}

The data underlying this article will be shared on reasonable request to the corresponding author.



\end{document}